\documentclass{edp-jp4}
\usepackage{graphicx}
\newcommand{\ltsima}{$\; \buildrel < \over \sim \;$}
\newcommand{\simlt}{\lower.5ex\hbox{\ltsima}}
\newcommand{\gtsima}{$\; \buildrel > \over \sim \;$}
\newcommand{\simgt}{\lower.5ex\hbox{\gtsima}}
\begin{document}

\title{M31, M33 and the Milky Way: Similarities and Differences} 
\author{Rosemary F.G.~Wyse}\address{The Johns Hopkins University, Department of Physics and Astronomy, Baltimore, MD 21218, USA}
\maketitle
\begin{abstract}
The large galaxies in the Local Group, while all disk galaxies, have
diverse stellar populations.  A better understanding of these
differences, and a physical understanding of the causes, requires more
detailed study of the older populations.  This presents a significant
challenge to GAIA but the scientific returns are also significant.
 \end{abstract}
%
\section{Introduction}
Study of the resolved stellar populations of galaxies in the Local
Group offers great scientific returns in our understanding of how
galaxies form and evolve. The Local Group member galaxies
(cf.~\cite{sid}) include the three disk galaxies M31, M33 and the
Milky Way, and numerous dwarf companions, both gas-rich and gas poor.
What causes their similarities and also their diversity?

There are two main aspects to galaxy formation and evolution, namely
the history of {\it mass assembly and re-arrangement\/} and the
history of {\it star formation}.  The old stellar populations play a
particular role in deciphering these histories, since old stars
usually retain a memory of certain aspects of their early life, such
as the surface chemical abundances, and often orbital angular momentum and orbital energy. 

The important questions concerning
the mass assembly, apart from its rate, both past and present-day,
include: what was the nature of the mass? -- since collisionless dark
matter, collisionless stars and collisional gas behave differently;
what was the density distribution of any and all components? -- since
physical processes such as tidal stripping depend on relative
densities, and dynamical friction timescale depends on mass ratios;
what are the specific angular momenta and orbits? -- since the
coupling efficiencies of various processes depend on these.

The important questions concerning star formation history include:
what was the rate of star formation and how did/does it vary as
function of spatial location?; what was and is the stellar Initial
Mass Function? -- the visibility of galaxies at high redshift,
their contributions to background light in different passbands, their 
chemical enrichment, stellar feedback, the supernova rate, gas
consumption rate etc.~depend on the IMF; what was/is the mode of star
formation? -- what fraction formed in super star clusters?; and of
course -- what is the connection to the history of mass assembly of the 
various components.

Spiral galaxies are clearly diverse in their properties, for example
in bulge-to-disk ratio, but theories should be able to produce the
galaxy population in the Local Group rather naturally, without appeal
to special conditions.  Thus the Local Group members are `typical'
galaxies in their properties and for theory, but they are atypical for
observation.  From their resolved stellar populations we can obtain
age distributions, chemical elemental abundance
distributions, kinematics -- all as a function of spatial location.
The tracers that can be used include stars of a range of evolutionary
stage and mass, planetary nebulae, star clusters, and gas through HII
regions, 21cm emission and CO emission. 
The stellar properties in satellite companion galaxies provide
important complementary constraints, for example, limiting the possible
contribution of disrupted satellites to larger systems (cf.~\cite{mukund}).

We have learned much about the Milky Way Galaxy from its resolved
stellar populations, but we still have only an incomplete picture, and
it is clear that GAIA will play a major role in furthering our
understanding.  Study of M31 and M33 with existing ground-based 
telescopes and with the Hubble Space Telescope has been limited, but as I
will describe below, has provided clear evidence of differences in
some aspects of their stellar populations and also of similarities. 

While detailed space motions and distances have a unique role to play
in understanding how the Local Group disk galaxies decompose into
different stellar components such as bulge/halo/disk/thick disk, much
can be inferred from mean kinematic quantities, such as the net
azimuthal streaming motion of a population.  To illustrate, Figure~1 shows the 
specific angular momentum distributions for these components of the Milky Way. 
The
similarity between the angular momentum curves for the bulge and
stellar halo can be explained by a model in which the proto-bulge gas
is ejected from star forming regions in the early halo \cite{rwgg92},
while the similarity between the curves for the thick and thin disks
is expected in a model where the thick disk is a remnant of the early
stages of disk formation (see below).  It is clear that ejecta from
the halo did not `pre-enrich' the disk.

\begin{figure}[!h]
\centering
\includegraphics[width=4in,angle=0]{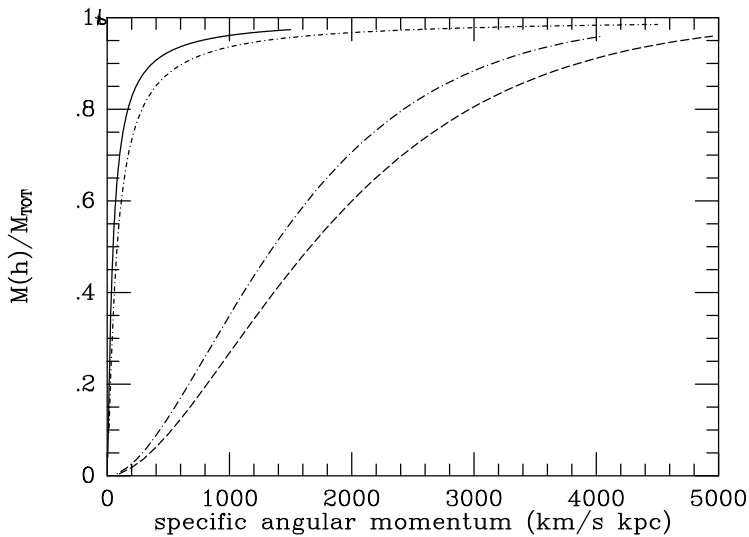}
\caption{Estimated angular momentum distributions for the stellar components of
the Milky Way.  The curves correspond to, from the left, the central
bulge, the stellar halo, the thick disk and the thin disk.  This
figure is taken from Wyse \& Gilmore \cite{rwgg92}.}

\end{figure}

Old stars are important for deciphering both the mass assembly and
star formation histories, and this poses a significant challenge for
GAIA capabilities, since at a distance modulus of $\sim 24.5$, the tip
of the Red Giant Branch (T-RGB) in M31 is at ${\rm I} \sim
20.5$\cite{durrell}, and there is no good evidence for an
intermediate-age population in the bulge of M31 that would contribute
Asymptotic Giant Branch (AGB) stars brighter than the
tip\cite{renzini}. However, significant scientific returns would be
achievable with mean kinematics of red giants in M31, M33 and their
satellites, as I describe below.  I hope to convey the exciting
science that could be possible were GAIA to push through its nominal
limit of  $G=20$ ($I \sim 20$), to reach below the T-RGB. Of course,
for those galaxies with (intermediate-age) stars brighter than the old
T-RGB, the nominal GAIA limit will still provide
significant results.

I will first describe some of the advances we have made so far through
study of the stellar populations of the Milky Way.  I will then
discuss what is known of the stellar populations of the remaining
large members of the Local Group, and raise some open questions,
including ones that may be addressed with GAIA.

\section{the Thick Disk and Constraints on 
Mass Assembly of the Milky Way Galaxy}

The effects of mergers between galaxies are to fatten disks, by
putting orbital energy of the galaxies into their internal degrees of
freedom, and to build up bulges and haloes, through a combination of
heating and angular momentum and mass re-arrangement resulting from
gravitational torques and bar formation, and assimilation of stars
removed from their parent galaxy by tidal effects.  The amplitudes of
these effects are dependent on the (many) parameters of the
interaction, such as mass ratio of the merging systems, gas content,
orbital inclination and angular momentum (both the sense and the
magnitude).  The age distributions, kinematics and metallicities of
the different stellar components of a galaxy -- and of different
tracers, such as young stars, old stars, or globular clusters -- are
very important in deciphering a complex situation, in which some
properties can be approximately conserved (such as angular momentum of
a stellar orbit or stellar metallicity) and some are not (such as the 
velocity dispersions of the disk stellar population).

A merger between a stellar disk and a stellar satellite of around
10-20\% by mass results in a fattening of the disk (as opposed to the
destruction of the disk that happens for mass ratios that are more
equal), and the thinness of disks can be used to constrain their
merging history (cf.~\cite{jpo}) and cosmological
parameters\cite{toth}. Indeed recent N-body
simulations\cite{vw,walker} have produced fattened disks that have
spatial distributions and kinematics rather similar to the thick disk
of the Milky Way Galaxy (see \cite{rome} for a recent review).  This
is particularly interesting for the mass assembly history, since the
thick disk, at least at the solar circle, is exclusively {\it old}, as
illustrated in Figure~2.  We know that the thin disk has been forming
stars fairly continuously over the last $\sim 12$~Gyr \cite{rocha},
with the consequence that if a significant merger had occured more
recently, younger stars would also be in a thicker disk. However,
these younger stars are not observed in the thick disk (cf.~\cite{carney,freeman,edvardsson,fuhrmann}). 
Thus the last
significant merger of the Milky Way occurred $\sim 12$~Gyr ago, when
the globular clusters like 47~Tuc were formed.  Of course, if the
thick disk is not the product of a `minor' merger, but e.g.~formed by
slow settling of the proto-disk to the disk plane \cite{hensler} then
the merging history is even more quiescent!

\begin{figure}[!h]
\centering
\includegraphics[width=3.25in,angle=270]{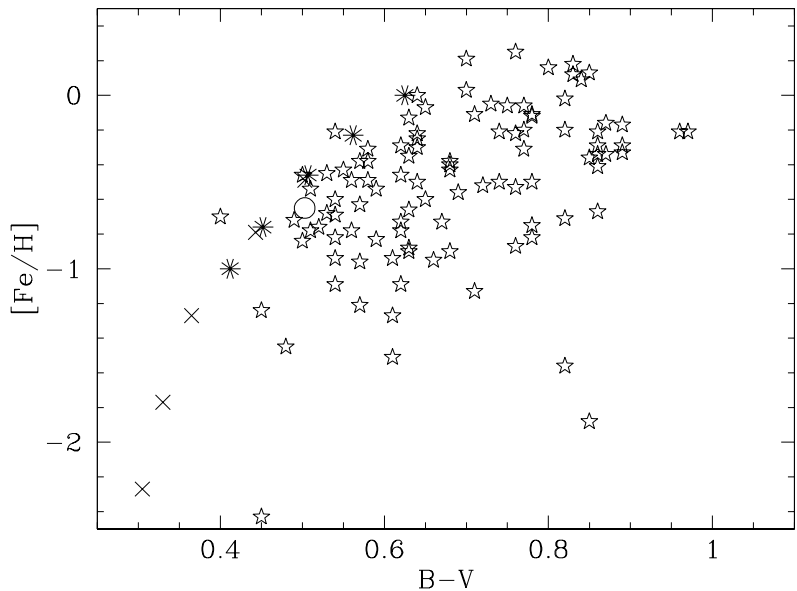}
\caption{Scatter plot of iron abundance {\it vs\/} B$-$V colour for
thick disk F/G stars, selected {\it in situ\/} in the South Galactic Pole 
at 1-2kpc above the
Galactic Plane (stars), together with the 14~Gyr turnoff colours
(crosses) from VandenBerg \& Bell\cite{vdbb} (Y=0.2) and 15~Gyr turnoff colours
(asterisks) from VandenBerg\cite{vdb} (Y=0.25).  The open circle represents the
turnoff colour (de-reddened) and metallicity of 47~Tuc (Hesser et
al.\cite{hesser}).  The vast majority of thick disk stars lie to the red of
these turnoff points, indicating that few, if any, stars in this
population are younger than this globular cluster.  This figure is
based on Fig.~6 of Gilmore, Wyse \& Jones\cite{gwj}.}

\end{figure}

Further, the thick disk contributes $\simgt 10$\% of stars in the
solar neighbourhood, and falls off perpendicular to the plane with a
scale-height 3--4 times that of the thin disk.  Thus a significant
fraction, some $\sim 30$\%, of disk stars at the solar Galactocentric
distance, 2--3 scale-lengths from the centre, formed at lookback times
of $\sim 12$Gyr, or at redshifts $\sim 2$.  This is not easily
understood in the context of hierarchical-clustering models of
structure formation, such as the cold-dark-matter scenario.  In these
models, the angular momentum transport that accompanies the merging
process to form galaxies \cite{navb,navs} leads to disks that
are too small, and one must appeal to `feedback' processes to delay
disk formation, to redshifts $\simlt$ unity\cite{weil,couchman}.
This is perhaps another way of saying that the merging history of the
Milky Way appears to be unusual in these models.  But is the Milky Way
unusual? What of the other large galaxies in the Local Group? There
are clearly some large, relaxed-looking disk galaxies at redshifts of
unity \cite{markd} -- can we identify their counterparts and descendants 
locally?

First, we need to establish the properties of the thick disk in the
Milky Way far from the solar Galactocentric radius; GAIA will play a
key role in this, providing accurate distances, metallicities, proper
motions and radial velocities for tracer stars (main sequence turn-off stars, red giants, Horizontal Branch etc.)  at distances as far as $\simlt
20$~kpc.

\section{The Thick Disk, Bulge or Halo of M31} 

The globular clusters in the Milky Way have a bimodal metallicity
distribution, with peaks at [Fe/H] $\sim -1.5$~dex and $\sim -0.6$~dex
\cite{zinn}, and around one-third being in the metal-rich population.
The clusters in these two peaks are further distinguished by
kinematics and spatial distribution.  The metal-poor population
consists of the classic halo globular clusters, which are old and on
orbits of low angular momentum.  The metallicity and kinematics
characteristic of the metal-rich globular clusters (of which 47~Tuc is
a member) are very similar to those of the thick disk, and these
clusters are usually ascribed to this population \cite{zinn,arm},
though their concentration towards the centre of the Galaxy has led to
their association with the bulge\cite{min,cote1}.  If indeed the last
significant merger event experienced by the Milky Way formed the thick
disk, one might expect there to be globular clusters associated with
it, by analogy with the young `globular clusters' (super star
clusters) identified in merging systems \cite{whit}.  Further, this
merger could have initiated bulge formation, consistent with the
similar characteristic ages of thick disk and bulge.

The globular clusters of M31 also have a bimodal metallicity
distribution \cite{barmby}, with peaks at metallicities remarkably
similar to those of the Milky Way system, and again most clusters
being in the metal-poor population. Furthermore these two populations
appear to be similarly distinct in their kinematics and spatial
distributions, as are the two Milky Way globular cluster populations.
The globular cluster systems of the Milky Way and M31 appear to be
analogues of each other.  Associating the more metal-rich globulars,
with peak [Fe/H] $ \sim -0.6$, with the thick disk, as in the Milky
Way, would lead to the expectation of a field thick disk with similar
mean metallicity.

The field stellar population of M31 has been studied through
colour-magnitude diagrams (both ground-based and from the Hubble Space
Telescope) of its evolved stars and spectroscopy of bright red giant
branch stars.  All studies
\cite{durrell,holland,rich96,raja,rich01,fergetal}, following the
pioneering work of Mould \& Kristian \cite{mould}, find that the
dominant field population probed down the minor axis has a mean
metallicity\footnote{Most of these metallicities are based on the
colour of the red giant branch and are subject to calibration
uncertainties including the elemental abundance mix.}  of around $\sim
-0.6$~dex, from projected distances of $\sim 5$~kpc out to $\sim
20$~kpc (the `bulge' minor-axis effective radius is $\sim 1.5$~kpc
\cite{rene}, significantly larger than that of the Milky way).  The
metallicity distribution is asymmetric, and can be fit by the
superposition of two populations, metal-poor and metal-rich, with
peaks similar to the globular clusters, at $\sim -1.5$~dex and $\sim
-0.6$~dex. The bulk of the stars, even out at 20~kpc, is in the
metal-rich population.  Thus unlike the case for the globular
clusters, the field stars of the `halo' in M31 are predominantly
metal-rich.  It should be remembered that these field stars are
members of Baade's `Population II' \cite{baade44,baade58}, raising the
issues of which stars in the Milky Way should have been identified as
`Pop II' -- perhaps \cite{rwgg88} the members of the Milky Way thick
disk, whose mean metallicity is comparable to that of the dominant
population in M31's `halo'.  Indeed, perhaps the `halo' in M31, which
is rather flattened with an axial ratio of $\sim 0.6$, is actually a
thick disk (cf.~\cite{rwgg88}).  This could have been formed during a
significant merger -- as discussed more fully in Ibata's contribution,
wide-area star counts of the evolved population in the `halo' of M31
have revealed a large overdensity most easily explained as a remnant
star stream from tidal interactions, albeit of the same high mean
metallicity.  Mean kinematics, as could be possible with GAIA, would
provide further signatures of `streams'.

The red giant luminosity function, in fields at projected distances
from the centre of $\sim 1$~kpc to $\sim 20$~kpc
\cite{steph,durrell}, is consistent with no young, luminous AGB stars,
providing a weak limit on the age distribution of the bulge/halo/thick
disk of M31, as older than a few Gyr.  The red giant luminosity
functions are in fact very similar to that of the evolved stars in the
Milky Way bulge (measured in Baade's window, $4^\circ$ from the
centre), for which we know from deep photometry below the turnoff that
the population is old \cite{feltz}.  Optical imaging of fields at
projected distances from the centre of $\sim 1$~kpc also show no
evidence for stars brighter than the tip of the RGB (after taking
careful account of blending; \cite{jab,renzini}).

The presence of RR Lyrae stars in the bulge/halo of M31 \cite{pv87}
argues for an old component, age $\simgt 10$~Gyr. Horizontal branch
morphology can provide clues, though is not yet available;
ground-based data thus far have prohibitively large errors by the $V
\simgt 25$ level of the HB. HST studies of M31 up to now have been
primarily globular cluster fields, mostly in fields with too much disk
contamination to study the field halo/bulge.  

Even more intriguing is the fact that colour-magnitude diagrams for
fields in lines-of-sight that should be predominantly outer disk have
a RGB morphology very similar to that of `halo'-dominated fields
\cite{ferg2001}, with an additional metal-rich component that may be
`thick disk' \cite{sara} or simply the thin disk.  There is apparently
a significant old component in these outer disk fields, with important
implications for the onset of disk formation \cite{ferg2001}, as
indicated above.  Or is the disk so warped that one cannot calculate
reliably its contribution in a given line-of-sight, based on simple
surface brightness profiles?

Thus the CMDs of M31 offer fascinating clues to the past history of
our nearest large galaxy, but kinematics and metallicities are
required to untangle the different populations projected into the same
line-of-sight.  Radial velocities of the bright giants are possible
with 10-m class telescopes \cite{raja}.  Multi-band wide-field mapping
of the field population below the tip of the RGB i.e.~I$\simgt 20.5$
is feasible with existing ground-based telescopes; GAIA could provide
the information necessary for their interpretation through
measurements of mean/systematic motions of populations defined by, for
example, colour or position (as a reminder, at the distance of
M31, $\sim 750$~kpc, expected individual proper motions are less than
$\sim 100\mu$arcsec/yr, somewhat less than the expected accuracy of
GAIA measurements for stars with $G=20$).  The surface brightness at
the effective radius of the `halo' of M31 is $\mu_B \sim 22$ mag/sq
arcsec \cite{rene}, and determining the limiting background stellar surface
brightness at which GAIA will achieve its full accuracy is obviously
important.  Out at 20kpc along the minor axis, where the bulge field
is still metal-rich, the surface brightness is only $\mu_V \sim
30$~mag/sq.~arcsec\cite{pdv94}. 

Even old, metal-rich stellar populations can contain small numbers of
stars brighter than the T-RGB, as evidenced by the handful of Long
Period Variables in the globular cluster 47~Tuc \cite{mont}.  These
are plausibly in the Thermally-Pulsing (TP) phase of the AGB
\cite{renziniaraa} and their increased luminosity relative to
metal-poor stellar populations may be related to mass loss at the
Helium shell flash that marks the onset of the TP-AGB.  Further,
non-variable bright AGB stars are expected as the descendents of any
`Blue Stragglers' that may have formed, either in the field itself or
perhaps in globular clusters that were later disrupted.
Identification of these rare stars, possible through the full coverage
across the face of the galaxy with GAIA, would be exciting.

\section{M33 -- halo/bulge and clusters}

The early work of Mould \& Kristian \cite{mould} established that the
field stars of M33  some $\sim 7$~kpc projected distance  from the centre of
that galaxy, along the minor axis and thus expected to have little contribution from the disk, have
a mean metallicity of only $\sim -2$~dex, with a small spread.  There
is a kinematic `halo' as traced by the globular clusters
\cite{schommer}.  Analysis of the CMDs resulting from deep imaging
with the Hubble Space telescope \cite{saram33} has shown that some of
these `halo' clusters have a red horizontal branch despite low
metallicity ($\sim -1.5$~dex), perhaps indicating a younger age ($\sim
7$~Gyr), with others probably as old as the classical Galactic halo
globular clusters.  The field surrounding the globular clusters
studied with HST are disk-dominated and show a complex star formation
history \cite{saram33}.

Luminous star clusters have been identified across the face of M33
from HST images \cite{chandar1,chandar2}, with ages (inferred from
integrated colours) ranging from $\simlt 10^7$~yr to the $\sim
10^{10}$~yr of classical halo globular clusters. Derived masses are in
the range of $10^2 M_\odot$ to $10^6 M_\odot$ and correlate with age,
but there are apparently clusters of masses greater than $\sim 10^4
M_\odot$ with ages of only a few hundred million years
\cite{chandar2}.  There are some similarities to the populous
intermediate-age clusters in the Large Magellanic Cloud \cite{elson}
-- is there some aspect of the star formation process in very
late-type disk galaxies that favours populous clusters?  The luminous
stars in these clusters should be easily accessible to GAIA, allowing
the association of the parent clusters with disk, thick disk, bulge or
halo. 

Ground-based near-IR (J,K) imaging with adaptive optics on the CFHT,
of the central regions of M33 (inner 18$^{\prime\prime}$, or $\sim
50$~pc) find the fascinating result \cite{davm33} that there is a
strong AGB component, indicative of a burst of star formation 1--3~Gyr
ago, but with metallicity only $\simlt -1$~dex, much lower than the
metallicity of the inner disk inferred from HII regions, of around the
solar value (e.g.~\cite{vil}).  Why is the M33 inner bulge so
different from that of M31 or of the Milky Way?  How far out does the AGB component extend? Kinematics of the AGB
population, and a larger-scale survey, are of obvious importance to
attempt to understand the connections between disk and `bulge', and
the metal-poor `halo'. These stars are brighter than the Tip-RGB (by
some 2.5 mag in the K-band, less so in the V-band) and should be
amenable for study with GAIA.

\section{M32 -- an elliptical since when?}

While not strictly part of my remit, the compact dwarf elliptical
galaxy M32 offers an intriguing target for GAIA.  The evolution of
this galaxy has clearly been strongly influenced by its proximity to
M31, and indeed a very plausible scenario invokes severe tidal
truncation \cite{wirth} perhaps of a former disk galaxy, to leave the
central bulge \cite{bekki}.  There is a ubiquitous bright AGB
population in M32, well-mixed with the underlying older stars
\cite{davm32}.  Integral-field spectroscopy of the inner regions
suggests that the typical population has around solar metallicity and
an age of $\sim 4$~Gyr \cite{delb}.  It has been speculated that the
bright AGB stars are the remnant of a merger of some smaller system
with M32 itself \cite{davm32}, or the result of gas inflow, star
formation and mixing during the stripping of a disk galaxy
\cite{bekki}. These bright stars are $\sim 3$~mag brighter than the
TRGB in the K-band, and again their mean kinematics should be 
measurable.  This information, especially if the lower luminosity
(older) stars are also accessible by pushing GAIA to its limits,
should allow us to distinguish these two possibilities.

The proper motion of the centre-of-mass of M32, which should be 
measurable with GAIA, is obviously very important for deciphering the
orbit and the past history of the interaction between M31 and M32.

\section{Summary}

Cosmic variance requires that we confront theories of galaxy formation
and evolution with the detailed properties of as diverse a sample of
galaxies as possible.  The Local Group offers the opportunity to study
large galaxies with similar large-scale morphologies but different
present-day stellar populations. GAIA could provide kinematic,
distance and metallicity information to aid the deciphering of the
histories of these different galaxies and thus the physics of galaxy
evolution.

\begin{acknowledgements}
I would like to thank the organisers for inviting me to this
stimulating school, and for their financial support. I really enjoyed
being back at Les Houches. 
\end{acknowledgements}

\end{document}